# Assessing Changes in Thinking about Troubleshooting in Physical Computing: A Clinical Interview Protocol with Failure Artifacts Scenarios

Luis Morales-Navarro, University of Pennsylvania, luismn@upenn.edu
Deborah A. Fields, Utah State University
Yasmin B. Kafai, University of Pennsylvania
Deepali Barapatre, University of Pennsylvania

**Abstract**

**Purpose:** The purpose of this paper is to examine how a clinical interview protocol with failure artifact scenarios can capture changes in high school students' explanations of troubleshooting processes in physical computing activities. We focus on physical computing since finding and fixing hardware and software bugs is a highly contextual practice that involves multiple interconnected domains and skills.

**Design/methodology/approach:** We developed and piloted a "failure artifact scenarios" clinical interview protocol. Youth were presented with buggy physical computing projects over video calls and asked for suggestions on how to fix them without having access to the actual project or its code. We applied this clinical interview protocol before and after an eight-week-long physical computing (more specifically, electronic textiles) unit. We analyzed matching pre- and post-interviews from 18 students at four different schools.

**Findings:** Our findings demonstrate how the protocol can capture change in students' thinking about troubleshooting by eliciting students' explanations of *specificity* of domain knowledge of problems, *multimodality* of physical computing, *iterative testing* of failure artifact scenarios, and *concreteness* of troubleshooting and problem solving processes.

**Originality:** Beyond tests and surveys used to assess debugging, which traditionally focus on correctness or student beliefs, our "failure artifact scenarios" clinical interview protocol reveals student troubleshooting-related thinking processes when encountering buggy projects. As an assessment tool, it may be useful to evaluate the change and development of students' abilities over time.

# Introduction

Debugging, a key computing practice (Grover and Pea, 2013; Lodi & Martini., 2021; Shute et al., 2017), is the process of finding errors and fixing them in computing systems (McCauley et al.,

2008; Michaeli & Romeike, 2021). Much research has demonstrated that students, especially programming novices, face multiple challenges when debugging code, ranging from identifying simple syntax problems to more complex semantic problems when programs run but do not function as intended (McCauley et al., 2008). These challenges are even more evident in physical computing applications such as robots or electronic textiles (e-textiles), where debugging needs to focus on both code and the physical hardware. This means that beyond common syntactic and semantic issues in their programs, students have to attend to electronic and other physical aspects of their designs (Desportes & Disalvo, 2019; Lui et al., 2024). For this reason, debugging in physical computing is a highly contextual practice, often also called by the broader term *troubleshooting* (Jonassen & Hung, 2006), a term which we will use in the remainder of this paper when referring to identifying and solving problems in physical computing. To better understand novices' learning to troubleshoot physical computing applications, we need instruments to capture and measure changes in how they engage with troubleshooting.

Yet there are few instruments available to assess changes in novice students' thinking about troubleshooting in K–12 contexts. Assessing students' understanding is a challenge since troubleshooting involves specific domain knowledge (Vessey, 1985), systemic processes or "global strategies" (Schaafstal & Schraagen, 2000), and emotional factors (DeLiema et al., 2023). Some approaches to assessing troubleshooting focus on very limited, context-specific domains, such as a designed bug embedded in code (Denny et al., 2020) or Parsons' problems where learners rearrange pieces of code to fix a bug (Ericson et al., 2022). Analytics and automated evaluation can often be applied to these types of assessments because they have singular, "correct" solutions. Similar to this approach but with an eye toward physical computing contexts, some surveys query students about domain-specific components and processes needed in common machinery or makerspace projects (Blikstein et al., 2017). In contrast, some have developed "common sense" scenarios that get at everyday troubleshooting processes, for instance, how to help someone fix a light that doesn't turn on (Simon et al., 2008) or escape rooms with carefully designed puzzles (Michaeli & Romeike, 2020). Complementing this list of more cognitive-focused assessments are surveys that capture more social-emotional components like grit, growth mindset, and learners' self-reported comfort levels with various aspects of physical computing (Scott & Guinea, 2013; Morales-Navarro et al., 2023). Each of these assessments has affordances related to time, thinking processes, and practices within specific domains. In their review of computational thinking assessments, Tang et al. (2020) outline several gaps, particularly in assessing students' computational thinking in high school (upper secondary) or higher grades, moving beyond just domain-specific programming skills for learning and surveys about social-emotional dispositions, developing assessments that could be used in informal settings, and utilizing interviews (such as think-aloud interviews) to capture more "complex mental operations" than traditional tests and surveys generally allow (p. 10).

In this paper, we report on the pilot of an exploratory instrument to appraise students' troubleshooting thought processes in the form of a clinical interview protocol for high school introductory computing students that focused on changes in students' thinking about "failure artifact scenarios" specific to electronic textiles (a type of physical computing projects). Electronic textiles (e-textiles hereafter) involve sewable, programmable circuits using conductive thread to connect sensors and actuators to a microcontroller (Buechley et al., 2013). We designed context-specific scenarios with problems in e-textile artifacts that were presented to students as projects that someone else had designed. In a way similar to Simon et al.'s (2008) instruments, but using e-textiles, we showed examples of "failure artifacts" that only partially worked. These were shared with participants over video conference as pictures along with the creators' intentions for the projects. For the second scenario, we made an interactive Scratch simulation of the project so that students could test different programmed conditions. Students provided advice on problems to look for and approaches to take in diagnosing and solving potential problems. The intentionally vague scenarios had multiple potential problems with no singular solution, since the goal was not to see "correct" solutions but rather to observe the thinking skills students applied when interacting with the scenarios. Analyzing 18 matching pre- and post-interviews for emergent themes, we asked the following research question: *What changes in students' explanations of troubleshooting processes does the failure artifact scenario protocol capture? In other words, in what observable ways did students' thinking about troubleshooting our designed failure scenarios change before and after the electronic textiles unit?*

Our findings illustrate ways in which the failure artifact scenarios elicited students' thinking about the *specificity* of domain knowledge of problems, *multimodality* of physical computing, *iterative testing* of failure artifact scenarios, and *concreteness* of troubleshooting and problem solving processes. In the discussion, we consider implications for developing tools to assess students' troubleshooting abilities through contextualized troubleshooting challenges in physical computing.

# Background

Troubleshooting involves both finding and fixing bugs; thus, in assessments, we must consider how learners construct a problem space, isolate or diagnose issues, and finally generate solutions (Jonassen, 2006; Michaeli & Romeike, 2021). Constructing a problem space requires learners to build a mental model to represent the system they are troubleshooting and how its parts are interconnected. Isolating issues or diagnosing them involves observing problems and ascribing plausible causes. Here, learners may rely on their experiences with issues they have encountered in the past (Konradt, 1995). Diagnosing issues often involves generating hypotheses and testing them at different levels; one may test the whole system or focus on only a part of it (Jonassen & Hung, 2006). Lastly, generating solutions involves coming up with plausible ways to address an identified issue. Further, taking a situated view of learning, where learning is a process of changing participation within communities of shared practices (Lave & Wenger, 1991), the context of troubleshooting matters, whether on a screen or part of a physical artifact, as part of an isolated exercise in a class or as a design task. Thus, capturing troubleshooting processes—from hypothesis generation to testing parts of the whole—needs to be done in a contextualized assessment.

Troubleshooting physical computing projects is a highly contextual practice because it requires understanding how specific hardware and software components interact (DesPortes & DiSalvo, 2019; Wagh, 2017), as well as how bugs can emerge across and within domains such as coding, circuitry, and other physical materials (Searle et al., 2018). Sensor-based projects present additional challenges because students must consider sensor inputs that do not always interact in smooth, predictable ways to trigger planned outputs such as lighting or audio patterns (Hennessy Elliot et al., 2023; Nixon et al., 2023; Jayathirtha, 2022; Lachney et al., 2021). Some of the challenges novices exhibit include identifying bug locations incorrectly (e.g., identify bugs in only one domain, such as circuitry), ignoring errors in one domain entirely (e.g., code), or incorrectly assuming bugs are solved (DesPortes & DiSalvo, 2019). While common struggles with debugging syntactic and semantic issues in code remain, Sadler and colleagues (2018) note that the most common errors in physical computing involve incorrectly connecting components (sensors and actuators) to the microcontroller. Indeed, physical computing newbies may make mistakes when building circuits, for example, connecting sensors or actuators to pins that do not have the right functionality or connecting pins in the wrong way when it comes to polarity (Alessandrini, 2023; Booth et al., 2022). Bugs in code involve syntax issues such as typos, using variables without declaring them, calling undefined functions, and missing delimiters. These are complemented by semantic bugs in code such as misplacing a delay, contradictory or redundant logical expressions in conditionals, and misuse of built-in functions (see DesPortes & DiSalvo, 2019, Lachney et al., 2021, Booth et al., 2022). As such, being able to identify plausible and specific bugs, generate multiple hypotheses, and understand how to troubleshoot across domains (i.e., circuitry and code) is essential for learning physical computing. Assessing this requires some acknowledgement of the complexity of the multi-domain nature of physical computing.

One approach to assessing processes of troubleshooting is to present learners with problems and observe how they diagnose and solve the issues. For instance, Michaeli and Romeike (2020) developed an instrument to assess everyday troubleshooting through escape room tasks. Within students' collaborative approaches to the set of tasks (which all had single solutions), they found that novices encountered problems formulating initial hypotheses, coming up with alternative hypotheses, or with multiple hypotheses of what the cause of a problem could be. This is not surprising since being able to generate plausible hypotheses and multiple hypotheses for a single issue is a documented difference between novice and experienced troubleshooters (Gugerty and Olson, 1986; Kim et al., 2018). However, the escape room task did not specifically cover the domain of physical computing, though it did require computational thinking. A somewhat different approach is to present students with a project embedded with multiple bugs; for instance, the "DebugIts" developed for studies of e-textiles (Fields et al., 2016; Jayarthirtha et al., 2024; Lui et al., 2024). Providing students with actual projects with multiple, ill-structured bugs in and across coding and circuitry allowed researchers to study students' thinking processes as they applied multiple strategies in problem-state representation, fault diagnosis, and solution generation and verification.

However, while the ill-structured problems of the escape room and the DebugIts allowed for the assessment of rich troubleshooting strategies, the time involved was lengthy (~45 minutes). Further, the studies developed for both the escape rooms and the e-textile DebugIts were only done at single time points, not allowing for the investigation of *changes* in students' troubleshooting processes. The approach to providing students with "failure artifact scenarios"—or scenarios of buggy physical computing projects—provides ill-structured problems that could involve multiple domains of physical computing. But instead of giving students an actual physical computing project and its code, which would require substantial time to investigate and solve, we provided students with *descriptions* of two buggy e-textiles projects that did not work as intended. Our goal was to create a time-limited (10-20 min), easily replicable task (i.e., onscreen), aligned with students' experiences and practices of designing e-textiles alongside peers, to capture changes in how students think about and advise approaching troubleshooting processes before and after a curricular unit. In the discussion, we consider possibilities for implementation in classroom settings.

# Methods

## Context

We scheduled our interviews before and after the e-textiles unit of the Exploring Computer Science (ECS) course, a 12-week, equity-focused, inquiry-based introductory computing course for secondary students that school districts all over the United States have adopted (Goode et al., 2012). The 10- to 12-week e-textiles unit (http://exploringcs.org/e-textiles) was designed for youth to create a series of personally relevant creative projects while learning new coding, circuitry, and crafting skills (Kafai et al., 2019). The unit requires students to apply computing concepts such as sequences, loops, conditionals, variables, nested conditionals, data input from sensors, and functions in a text-based programming language (Arduino). Included in this study between the third and fourth projects of the e-textiles unit was an additional 7-day debugging activity, where students created intentionally buggy projects for their peers to solve (for more on the activity, see Fields et al., 2021; Morales-Navarro et al., 2024). In total, students spent 8–12 weeks creating, coding, and debugging e-textile projects.

Of note, because all schools in the area had virtual schooling due to COVID-19 in Spring 2021, all activities of this study—learning and research—took place virtually through Zoom. Students were provided materials through pick-up at school or directly through the mail, depending on each school's circumstances. The institutional review board at the University of Pennsylvania approved the study protocol.

In Spring 2021, four experienced teachers at different secondary schools in the Western United States, with high percentages of historically marginalized in computing secondary student populations (58–95% free and reduced lunch; 85–99% non-white), taught eight ECS classes with e-textiles. Two researchers who participated in the classes conducted pre-interviews with 33 consenting students. Students were selected randomly from those who provided informed consent (parents) and assent (students), with ~4-5 students per class. Because of the timing of the study at the end of the virtual school year in the second year of the pandemic, there was a high degree of attrition in student participation; many students were no longer participating by the end of the school year. In the end, we had 18 matching pre- and post-interviews.

## Data Collection

Data were collected through semi-structured clinical interviews. During clinical interviews, students are presented with problem situations and asked to solve, explain, or reflect on them (DiSessa, 2007). Interviewers encourage students to verbalize and explain their thinking processes in order to uncover student understanding (ibid). These types of interviews can also be used as assessment tools by teachers and researchers to better understand students' knowledge and experiences in specific domains, allowing them to tailor their instruction to build on their current knowledge (Russ & Sherin, 2013). In computing, clinical interviews with undergraduate students have been used to investigate how they construct knowledge around difficult concepts in CS1 (Yuen, 2007), as well as to assess how learners think about how commercially available physical computing artifacts work after instruction (Lee & Fields, 2017). In K-12 computing education, interviews have occasionally been used to assess programming abilities (see Sentance et al., 2023), allowing researchers to prompt learners to demonstrate their understanding of concepts (Grover et al., 2015) and interact with design scenarios to fix bugs and remix projects (Brennan & Resnick, 2012).

In this study, we created a clinical interview protocol based on the tradition of assigning students a buggy project to solve: a failure artifact scenario. However, rather than providing students with an actual physical computing project and its code, which would take significant time to investigate and resolve, we provided students with descriptions of two buggy e-textiles projects that did not function as intended. As such, interviews involved a set of two designed failure artifact scenarios with prompts intended to solicit students' explanations and reasoning in an open-ended manner, i.e.. We intentionally designed these interviews to be similar to those used in conceptual change research (diSessa et al, 2007; Lee & Fields, 2017; Sherin et al., 2012). The appeal of open-ended conceptual change-style interviewing was that it could be accessible to novices, elicit a broad range of student ideas (e.g., about troubleshooting, e-textiles, and computing), and follow up with questioning for increased robustness. We designed the two failure artifact scenarios to include common potential bugs or problems that students may encounter in an introductory electronic textiles unit (e.g., short or open circuits, not declaring a variable, embedded on fabric textile projects) (Lui et al., 2024). Further, aligning the scenarios with students' practices of learning physical computing in the context of designing personally relevant artifacts in the electronic textiles unit, we framed the scenarios as designs created by peers (or the interviewer) that were not working as intended. This aligns with common practices of giving and seeking peer support in classroom practice (Fields et al., 2021). Interviews took an average of 10.5 minutes each, lasting between five and 25 minutes.

The interview protocol we developed (see Appendix 1) invited students to make suggestions for how to troubleshoot two projects, called Broken Bird and Captain America, that someone else had made. Images of the buggy projects and descriptions of how these were supposed to work are shown in Figure 1 (right and left). The interviewer read the descriptions out loud and made a written description available to students, then asked students for help: "What would you tell this student to do in order to fix the project?" We designed this question to elicit students' current troubleshooting practices—to learn what approaches, hypotheses, and ideas students had about how to fix a broken project.

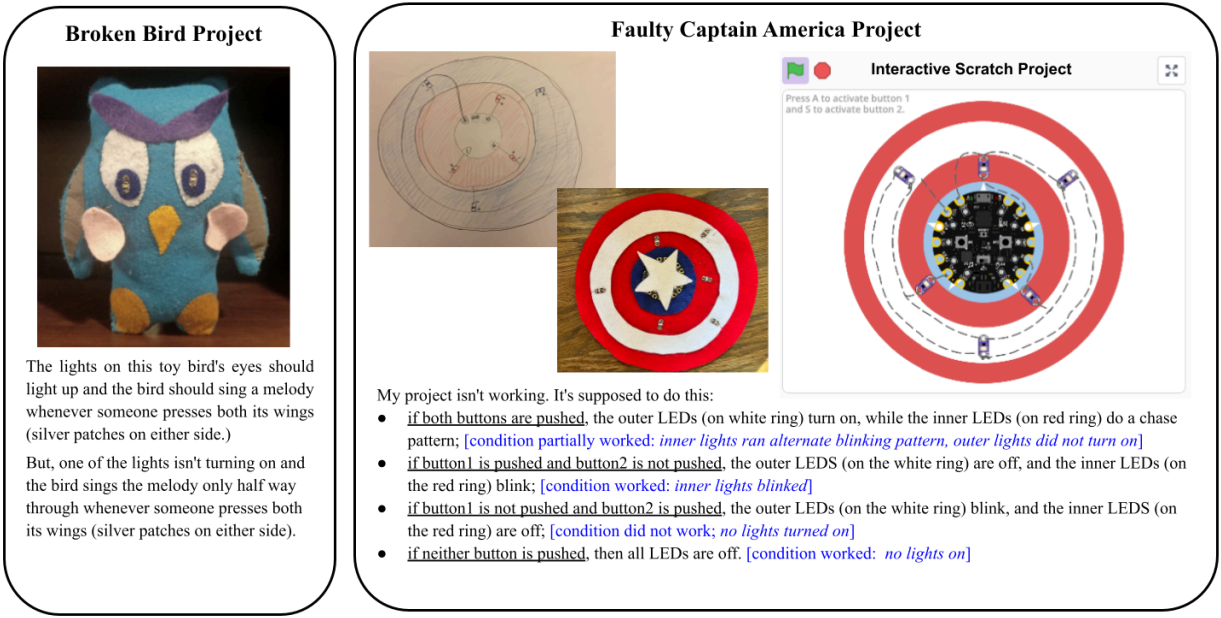

**Figure 1.** Interview Prompts Featuring Broken Projects. Comments in blue indicate actual functionality of interactive Scratch project

For the second project, a Captain America shield, (see Figure 1, right) we additionally presented students with an interactive Scratch project (https://scratch.mit.edu/projects/479329018) that enabled them to test the project and its functionality. If students carefully tested all four possible conditions with two switches, they would notice that the outer lights did not work under any condition, but that pressing button2 and button1 at the same time resulted in a different pattern for the inner lights than just pressing button1. This was important since it meant that nothing was wrong with button2 as a mechanism; pressing it triggered a new condition. Thus, the second scenario became a means of assessing how systematically students tested the project and whether they could eliminate some possible problems through that testing. In the post-interviews, the scenarios remained almost identical but with slightly different-looking projects (e.g., an elephant and a teddy bear with identical circuitry and functioning). This was intentional: to present students with the equivalent problems in both pre- and post- with scenarios that looked aesthetically different.

For both projects, when students ran out of ideas, we used follow-up prompts: "What do you think could be the causes of each of these issues? How would you fix them?" This allowed us to note whether students' ideas came up without or with prompting.

In sum, we collected 18 pre- and post-interviews in two rounds, which resulted in a total of four hours and 51 minutes of video recordings. Interviews were transcribed, and transcriptions were checked for accuracy.

# Analysis

In our analysis, we build on traditions of qualitative and learning sciences research in computer science education research (Tenenberg et al., 2019; Margulieux et al., 2019) in inductively studying students' troubleshooting practices as exhibited in their think-aloud explanations and observable activities (e.g., testing). During the first round of analysis, two researchers inductively coded a third of the data (6 sets of matching pre-/post-interviews) to develop an initial coding scheme. Researchers watched the video recording simultaneously while analyzing the transcripts. This was particularly important to be able to capture non-verbal actions of students, such as how they annotated a diagram or how they tested all the possible conditions in scenario 2. Instead of creating separate video logs, we added notes to transcripts on how students tested and annotated the projects. We checked and revised this scheme several times in group meetings and with application to additional interviews, then created a codebook with categories for students' articulation of possible bugs, multiple causes, next steps for fixing bugs, debugging process, testing, and whether students made references to their personal experiences in relation to debugging. Each of these categories included codes with details about the domain in which these applied and whether actions or discourse were prompted by researchers. For example, the category "articulating bugs" had three codes for how students talked about identifiying possible code and circuit bugs with subcodes that described how specific the bugs were (see coding scheme in Appendix 2. for more details). While these code categories were developed inductively, they relate to common aspects of troubleshooting (such as considering multiple causes) that are challenging for novices (Michaili and Romeike, 2022) as well as physical computing-specific challenges (such as identifying bugs across domains) (Lui et al., 2024). In a second round of analysis, two researchers applied the coding scheme across all pre-/post-interviews, co-watching the video recordings together while also annotating the transcripts to capture gestures and participant testing of the interactive simulation of the broken project. While coding together, the researchers dialogically engaged with the data, seeking agreement and iteratively discussing disagreements with a third researcher familiar with the data and the coding scheme. Since this is an exploratory study with a small number of participants, we prioritized establishing unanimous agreement on all coding (by coding together and reconciling our different opinions through extensive discussion) over reliability (when coders apply the same scheme independently to the same data) (McDonald et al., 2019). To visualize changes in students' explanations and thinking from the pre-interview to the post-interview across all participants, we created alluvial diagrams. These kinds of diagrams are useful to depict change over time when individuals are assigned to groups. As such, we agregated the codes and assigned groups to students (e.g., we assigned different groups to students that articulated code and circuit bugs, articulated only circuit bugs, only coding bugs, or no bugs at all). With the diagrams, we were able to visualize how the distribution of students by groups changed from pre- to post-. All names used in the paper are pseudonyms.

# Findings

Prompting students with failure artifact scenarios in clinical interviews enabled us to observe not only changes but also some growth in thinking about troubleshooting e-textiles. In particular, between pre- and post-interviews, we identified development in students' thinking across domains of physical computing in articulating possible bugs, articulating multiple potential causes for bugs, and being more specific in articulating possible bugs. We also saw evidence of change in students' suggestions to test projects and troubleshoot processes to apply in diagnosing problems, as well as the concrete next steps they articulated to solve problems.

## Capturing thinking across domains of physical computing

The failure artifact scenario instrument enabled us to observe how, from pre- to post-, students improved their articulation of possible bugs across multiple domains: code, circuitry, and both domains (see Figure 2). Considering multiple domains when articulating possible bugs involves constructing a complex problem space that involves both code, circuitry, and the interactions between domains. Whereas prior to the intervention, eight students (44.44%) talked about both possible circuit and coding bugs; afterwards, 13 students (72.22%) talked about potential bugs in both domains. After the intervention, students were aware that bugs in physical computing systems may be present both in circuitry and code. As an example, in contrast to her pre-interview, in her post-interview, Umar suggested multiple sources of problems—in the stitching, in the code, or even in a mismatch between circuitry connections and named pins in the code. This covers multiple domains of physical computing in her troubleshooting ideas.

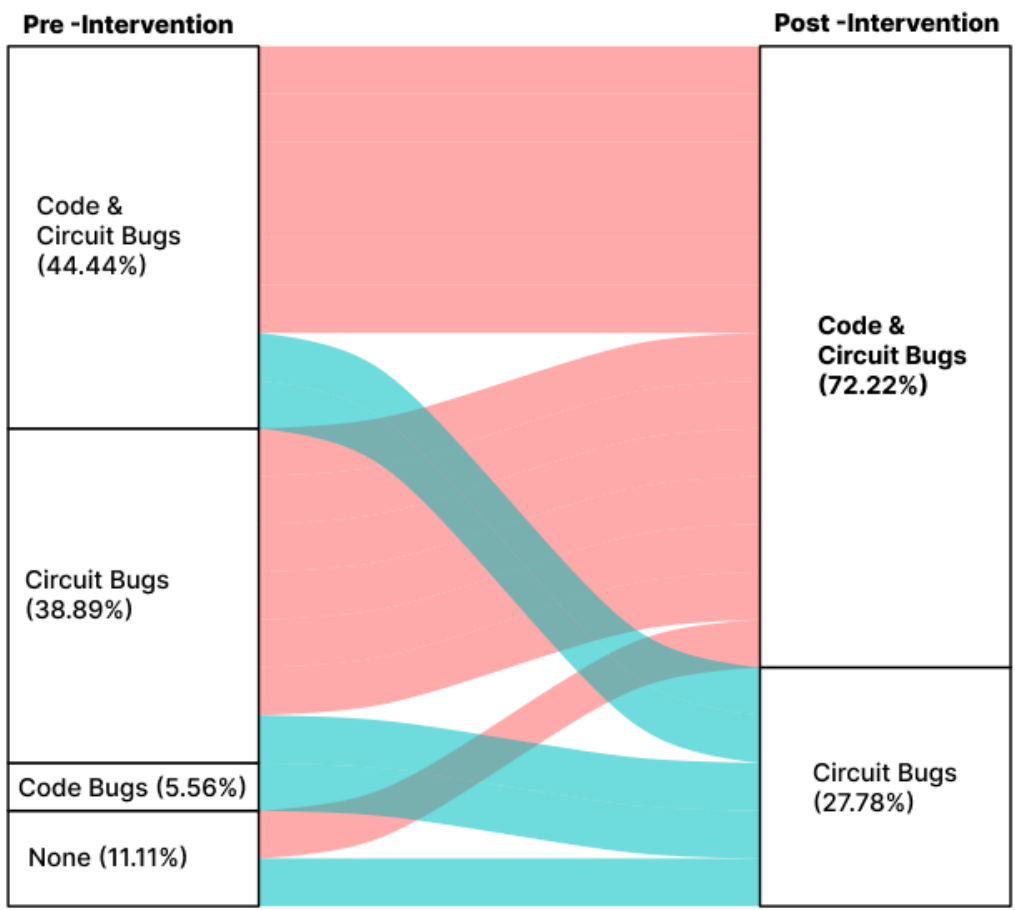

**Figure 2:** An alluvial diagram illustrates the changing distribution of students' articulation of bugs across multiple domains.

## Changes in thinking about multiple potential causes for bugs

Second, the instrument captured pronounced differences between pre- and post- in students articulating *multiple potential causes* for bugs within single domains and across multiple domains (e.g., circuitry and code, see Figure 3). For instance, an LED light could fail to turn on because of a short circuit or because the pin that connects it to the microcontroller was not correctly declared in the code. Overall, while in pre- only two students (12%) voiced that a bug could be caused by more than one issue, in post- nine students (50%) explained that a bug may have multiple causes. For example, Ava went from not suggesting any possible bugs in pre- to explaining that a light may not turn on because of two possible causes across domains: a faulty connection or incorrectly declaring a variable that stores the number of the pin where the LED light is connected.

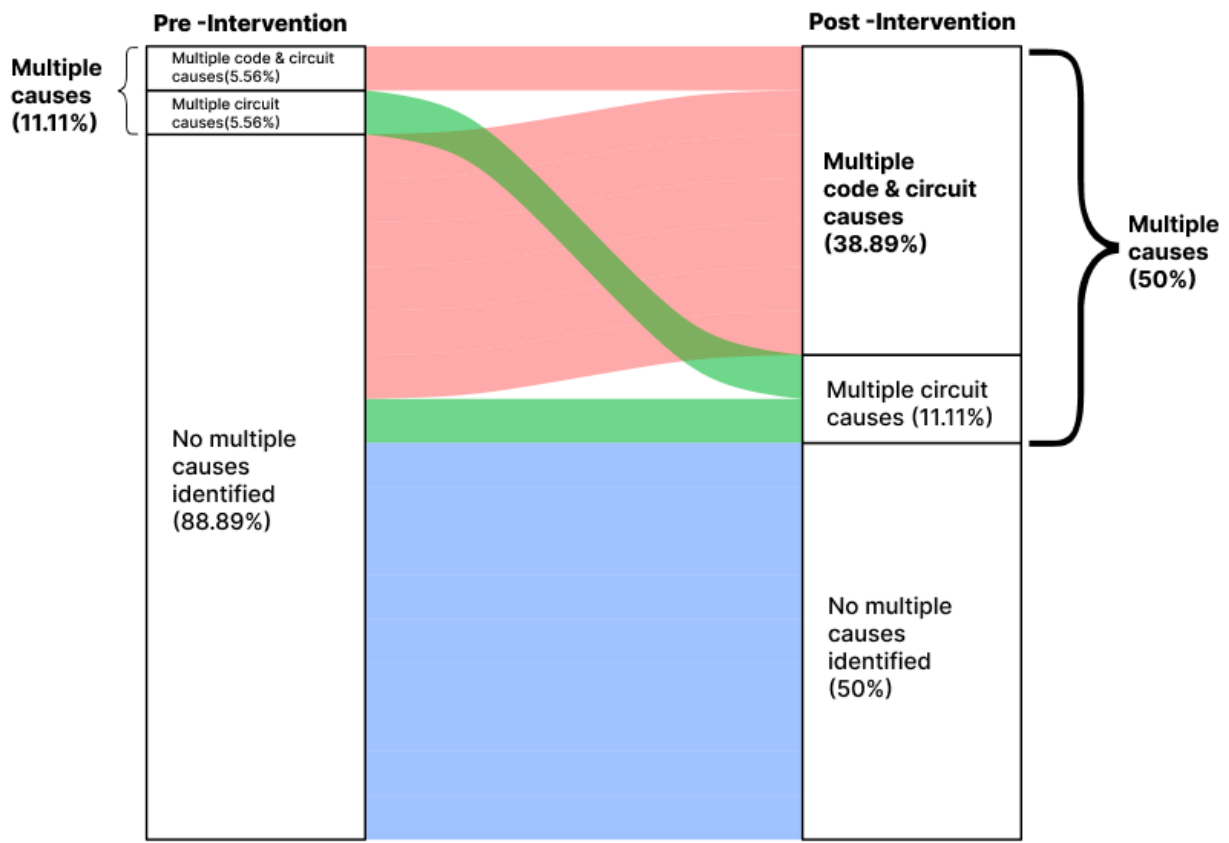

**Figure 3**. Alluvial diagram showing changes in the distribution of students' articulation of multiple causes for bugs.

Further, only two students in pre- articulated that a potential bug may be caused by multiple issues (and only one explained that bugs could be caused by multiple issues across domains), while in post, seven (38%) students articulated that individual potential bugs—such as lights not turning on, a motor not moving, a button not working, or lights not blinking—could be caused by multiple issues across domains in coding or circuitry. For example, Damian explained that an LED could fail because the creator of the project did not declare the variable for the pin correctly, because they did not pass the right arguments to the digitalWrite() function, or because of a loose connection to the circuit board. This follows computing education literature noting that novices struggle to generate multiple hypotheses for bugs (e.g., Gugerty, 1986), something that must also cover multiple domains in contexts of physical computing (Lui et al, 2024).

## Changes in thinking about the specificity of potential bugs

The instrument also enabled us to observe changes in the *specificity* of the possible bugs that students talked about. We classified the potential bugs students articulated into three categories—defined, semi-defined, and undefined bugs—in domains of coding as well as circuitry/crafting (see Table 1).

**Table 1.** Levels of bug definition in coding and circuitry/crafting.

| Bug Articulation Categories | Talks about coding or circuitry/crafting as the relevant domain with varying levels of definition | Example |
|---|---|---|
| Defined Potential Bugs | Describes a bug location and articulates a specific error. | Coding: "Using digitaRead*() instead of digitalWrite() to change that LED to HIGH. - Damien |
| | | Circuitry/Crafting: "Is it because [the thread]'s not connected to GND?"- Noah |
| Semi-defined Potential Bugs | Describes either a bug location or a specific error, but not both | Coding: "There's something wrong in the for loop" ~Viviana |
| | | Circuitry/Crafting: "See if your knots are tighten[ed] or cut correctly," - Charlie |
| Undefined Potential Bugs | No bug locations or specific errors articulated | Coding: "Since there's two it's maybe one of the lights isn't coded in yet" - Amelia |
| | | Circuitry/Crafting: "something is wrong with the cables inside of the bird" - Cameron |

## Changes in thinking about the specificity of potential coding bugs

In addition to articulating more plausible coding bugs in post, the responses to the failure artifact scenarios also showed that students more clearly defined possible coding bugs after the e-textiles unit (from 9 students (50%) in pre- to 13 students (72.22%) in post). The number of possible bugs articulated by students increased from pre- to post-, with 1.89 bugs on average (SD = 1.05) in the pre-interview and 3.85 bugs on average (SD = 3.11) in the post-interview. Bugs included both semantic (e.g., flawed logic in a conditional statement) and syntactic (e.g., missing a delimiter) issues.

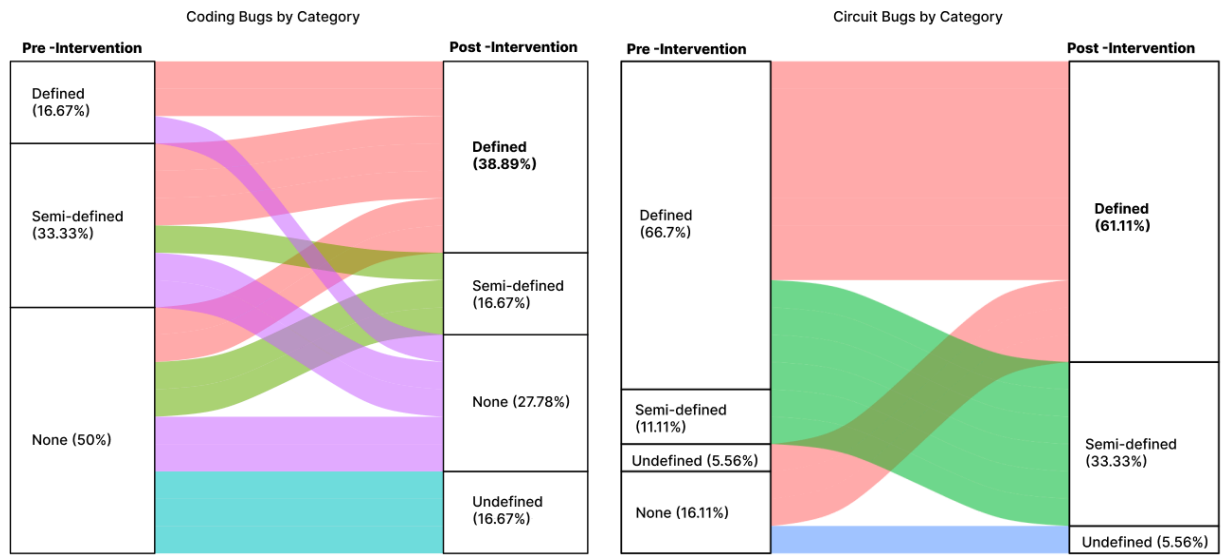

**Figure 4.** Left: Alluvial diagram showing student distribution by articulation of possible coding bugs. Right: Alluvial diagram showing how student distribution by articulation of possible circuit bugs changed.

Observing the level of detail at which students came up with possible bugs captured qualitative changes in how they diagnosed bugs and generated hypotheses. Explanations that included undefined bugs pointed to coding errors without providing any details on what the bugs could be or where they could be located: "maybe something's wrong with the code" or "maybe in the code, she didn't write it right, or she like missed a step, that's why one of the LED light is not on". These vague explanations acknowledge coding as a problem domain, but without much clarity.

In contrast, naming semi-defined bugs included slightly more detail in what and where a bug might be: "maybe you have to fix like an input" of a button or "there's something wrong in the for loop". These instances show some understanding of where in the code to look for bugs without specific ideas of what the bugs could be. Defined bugs showed further detail, for instance checking a variable used to store the value of data inputs from switches or suggesting a flipped error of input readings (in Arduino these are "HIGH" or "LOW"). Studying students' level of definition in articulating coding bugs was a productive way to observe changes in troubleshooting.

## Changes in thinking about the specificity of possible circuitry bugs

Students also changed in their level of specificity for articulating possible circuitry bugs, but in ways that require more nuance to understand. While all students talked about possible circuitry bugs in post- (15 students in pre- (83.33%), 18 students (100%) in post, see Figure 4 (right)), the level of definition of circuitry bugs shifted in ways that actually suggest deeper sophistication. As Figure 4 (right) shows, more students overall articulated possible defined and semi-defined circuitry bugs, though there were slightly fewer students in the category of highest definition (from 12 to 11). Yet looking at the actual explanations, some students shifted from defined circuitry bugs less likely to occur in practice (e.g., broken LEDs or button switches) to semi-defined circuitry bugs more likely to occur (e.g., loose or overlapping threads in vague locations). Whereas broken or unconnected parts are possible (and certainly well defined), it's more likely that loose connections, short circuits (overlapping threads), or reverse polarity issues would cause problems in e-textiles. This points to the importance of domain knowledge and experience in mapping the problem space (Vessey, 1985).

## Changes in systematically testing projects

The instrument also caught changes in students' systematically testing the projects for the second scenario presented in the interviews. Recall that in this scenario, students were given a simulation in which they could actively test a project and observe if its behavior matched the desired behavior. Fully testing the project involved students pressing four different combinations of two buttons to see if the light patterns matched those that were expected. While in pre- 10 students (55.56%) tested the project, in post- 15 tested them (83.33%) (see Figure 5). Additionally, we made a distinction between whether students tested on their own or after researcher prompting (see "Data Collection"). From pre-to-post, the number of students that tested without prompting increased from 8 to 12. For instance, in pre-interview, after researcher prompting to "press to activate the buttons," Cameron noted that "when you press [button] S none of the outer lights are working." In the post, he immediately started testing button combinations while voicing the behaviors he observed: "so I can press button one; it doesn't blink; button two works. Button one does not, for some reason, the eyes are not blinking." Being able to capture students' testing practices with an instrument is important, as testing can help students come up with hypotheses, test hypotheses, and isolate and decompose problems (Jonassen & Hung, 2006).

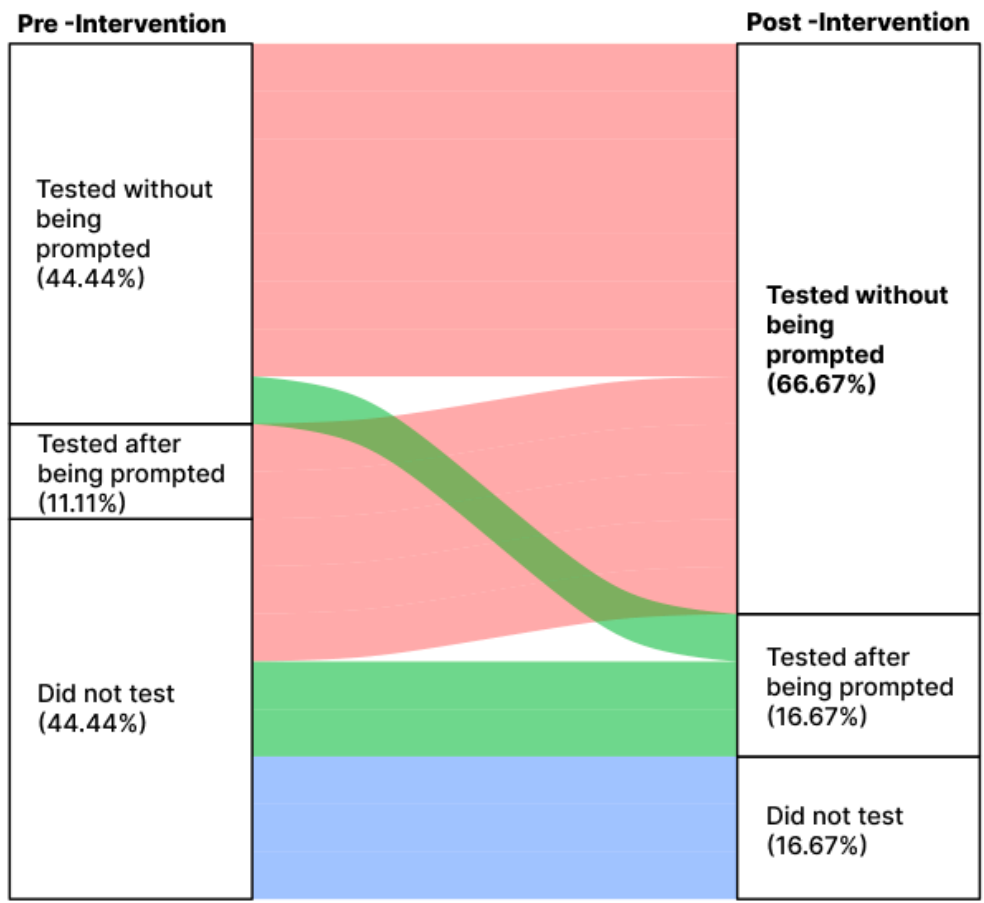

**Figure 5.** Alluvial diagram showing changes in student distribution by testing category.

## Articulating next steps for troubleshooting

The instrument also enabled us to assess how students thought about next steps for troubleshooting. We classified students' ideas into two categories: concrete and vague (see Figure 6). Vague steps involve a relevant domain but not a specific location (e.g., a component in the circuit or section of the code). For instance, Matthew vaguely suggested that the designer "just double check [the code] basically and test it to see what's wrong." Next steps became more specific in both detailing the location and actions that should be taken for troubleshooting within a relevant domain. As an example, Fernando suggested, "maybe connect this one to this one [gesturing to two LEDs]" to troubleshoot a light that was not turning on, proposing a specific step to isolate a problem. In both pre- and post-interviews 10 (55.56%) students voiced next steps for troubleshooting, yet the number of students that proposed concrete next steps increased slightly from five (22.22%) to seven (38.89%). This suggests that the specificity of the proposed next steps is helpful in differentiating students' troubleshooting.

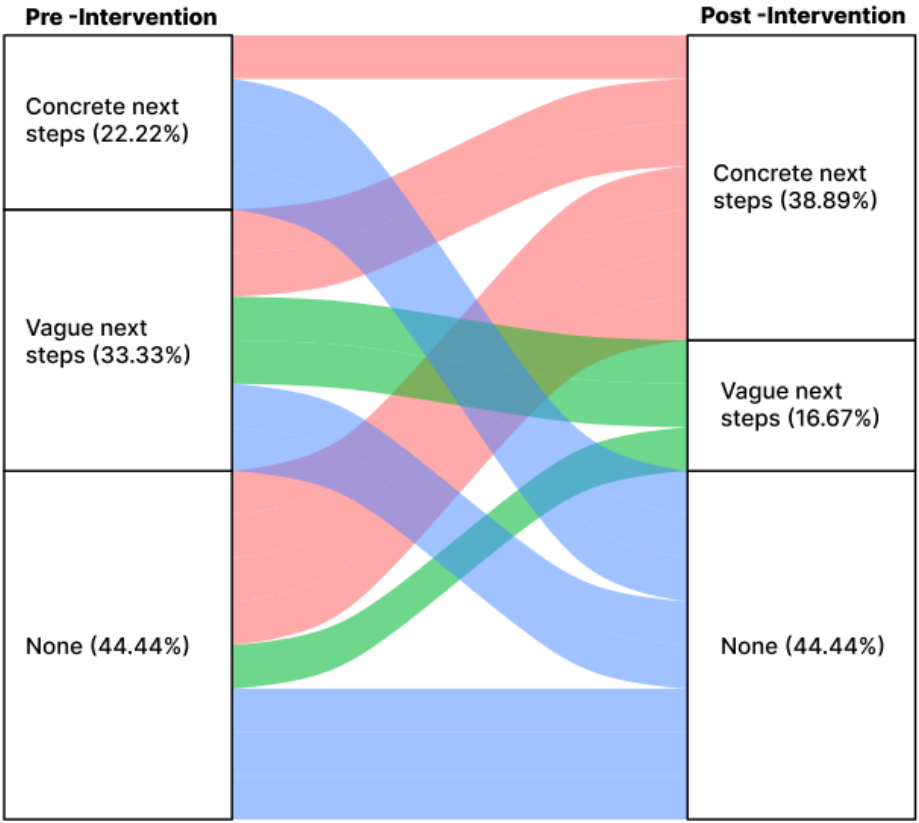

**Figure 6.** Alluvial diagram showing changes in student distribution by next-step categories.

## Differences in Students' Recommended Processes for Solving Bugs

In considering the failure artifacts, students often suggested ways to solve bugs, and we identified whether they suggested solving processes across domains (circuitry and coding) as well as whether the suggested processes were vague or defined. Interacting with the projects, students articulated sequences of steps for identifying and solving bugs across domains and with different levels of definition (see Figure 7). Defined problem-solving processes included recommendations to scrutinize circuitry such as Noah's suggestion to "check if one of your LEDS is upside down or try to see if you messed up any of the symbols, like the negative and the positive," or to follow a sequence of steps in looking at code, for instance Damian's recommendation to run the compiler to look for "any runtime errors" before looking for other mistakes, such as using the wrong variable. Four students in pre- and four in post- voiced sequences of steps that involving both circuit and coding bugs, including Ava, who voiced that to solve the problem of an LED not turning on, "you need to check your wiring for that specific LED light and see if it is in the right place and if everything looks good" before you "double check your coding to see if there are any slight errors."

Interestingly, fewer students suggested processes to fix bugs in post- (from 17 in pre- to 12 in post, see Figure 7), although the number of students who suggested processes that involved coding increased (from two to four), and number of students suggesting undefined processes (e.g., looking at the circuit and fixing it or "double check your work") decreased (from five to one). Some of these same students demonstrated greater sophistication in other areas, such as identifying possible defined bugs and testing the projects systematically. This shows that the development of troubleshooting thinking processes is not the same for all students and that growth in some aspects of troubleshooting such as identifying possible bugs may be more visible in a novice than being able to articulate a comprehensive strategy for troubleshooting.

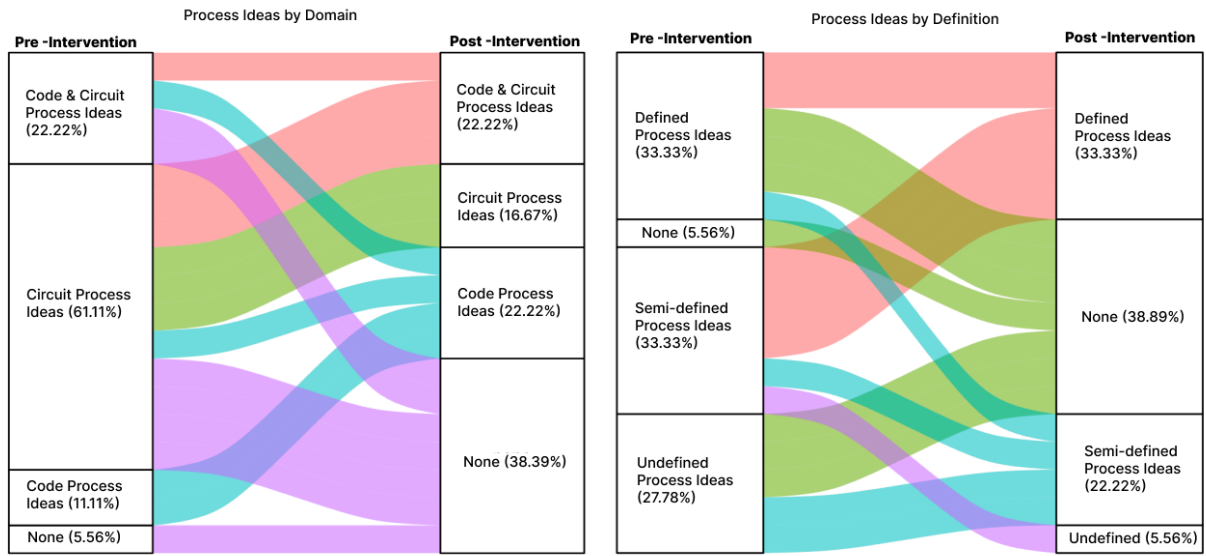

**Figure 7.** Left: Alluvial diagram showing changes in student distribution by domains of suggested processes for solving bug categories. Right: Alluvial diagram showing changes in student distribution by definition of suggested processes for solving bug categories.

# Discussion

Our goal in this study was to identify changes in students' troubleshooting thinking before and after a curricular unit on electronic textiles, a physical computing activity. We studied this through clinical interviews of students giving ideas of how to problem solve e-textile projects described to them (i.e., failure artifact scenarios). Our design and study of failure artifact scenarios provided a means of assessing changes in novice students' troubleshooting thinking processes. Targeted to students in high school (grades 9-12), moving to contexts beyond just screen-based computing as well as single-error assessments, it explores a way to fill some noted gaps called out by Tang and colleagues (2020) in their review, demonstrating one way to study students' "complex mental operations" (p. 10) around debugging and troubleshooting. Students' thinking about troubleshooting in physical computing requires understanding how different hardware and software components may interact, as well as how bugs could emerge across and within different domains. Failure artifact scenarios have relevance beyond the physical computing context of electronic textiles presented in this paper. There are many other examples where students already design fully functional digital media such as games, animations, and stories and where their learning could be assessed by presenting them with "faulty" versions or failure artifact scenarios to better understand their thinking and learning.

Thus, what we captured in the particular context of electronic textiles has potential for other assessing student learning in other contexts in which we ask students to create and share artifacts. In the following sections, we discuss in more detail what these findings mean in terms of understanding students' growth and developing tools for assessing their troubleshooting.

## Capturing Students' Changes and Growth in Thinking about Troubleshooting

The changes in students' suggestions for troubleshooting identified through our ground-up pre-/post- interview comparisons include: thinking across domains of physical computing; articulating multiple potential causes for bugs; articulating the specificity of bugs; testing projects in a systematic way; describing next steps for troubleshooting; and thinking about the process of debugging. Each of these changes is important and part of the larger suite of complexity in troubleshooting physical computing projects. Being able to capture how students think across domains and articulate bugs across domains and at different levels of specificity is particularly important since physical computing novices often identify bug locations incorrectly, assuming errors occur mostly in circuitry (DesPortes, 2019). At the same time, observing students' capacity to articulate multiple causes for potential bugs across domains is key, as this is a documented difference between novice and more experienced debuggers (Kim, 2018; Michaeli, 2020). Being able to capture students' systematic testing and the development of systematic strategies for troubleshooting is also crucial, as these are key aspects of troubleshooting that novices struggle with (Böttcher et al., 2016; McCauley et al., 2008). These aspects capture students' development in understanding of the problem space, as well as the repertoire of previous troubleshooting experiences they drew upon.

Of note, the range of changes in thinking we observed does not presuppose linear development of troubleshooting expertise. In general, shifting from articulating singular to multiple domains of bug sources and solutions, from vague to more specific types of bugs, and from vague to clear steps for further problem diagnosis and solving may be productive trajectories. However, we must treat students' changes with some nuance. For instance, one sign of growth in some students was shifting away from articulating specific circuitry bugs such as power sources or faulty parts (i.e., replace the LED, switch, microcontroller, or battery), demonstrating that concepts such as specificity need to be treated carefully as demonstrations of sophistication of troubleshooting. Further analysis could study individual trajectories of growth in troubleshooting physical computing. For instance, spider or radar charts with points at the different aspects of analysis in this paper (perhaps with other aspects of troubleshooting) might show gradations of learning of different individuals (see Lee and Fields, 2017). Additional analysis could delve into the quality or likelihood of bugs proposed by students. Further analysis of how proposed bugs may reflect student expertise could be helpful to have a more nuanced understanding of the types of bugs they diagnose.

# Developing Tools for Understanding Students' Troubleshooting

In contrast to simple exercises that focus on singular bugs, the use of failure artifact scenarios allows an investigation of students' troubleshooting thinking that is authentic to novice students' physical computing designs. Thus, these types of scenarios may provide tools for teachers (in classrooms) or, more broadly, educators in informal settings such as libraries, after-school clubs, and makerspaces to assess students' learning by focusing on students' thinking about troubleshooting (e.g., Koren et al., 2023). While being able to design functional circuits or write code are critical, creating functional physical computing artifacts lies at the intersection of these two domains. As Russ and Sherin (2013) demonstrate, interviews like the one in this study can serve as assessment tools for teachers to better understand students' knowledge. Failure artifact scenarios also push learners to think beyond the specifics of their own projects to consider problems in others' projects as a form of near-transfer, where the problem is slightly removed from the specific learners' creations (Klahr & Caver, 1988; Kwon & Jonassen, 2011). Future design and research should investigate how educators may adopt failure artifact scenarios in a variety of learning contexts.

Further, with the introduction of failure artifact scenarios in clinical interviews to assess troubleshooting, we present physical computing educators with an approach to engage students, individually or collaboratively, in demonstrating their newly gained competencies. In our particular case, we designed the failure artifact scenarios based on our extensive knowledge of the challenges novices encounter when creating physical computing artifacts (Liu et al. 2024; Jayathirtha et al., 2024). The problems were explicitly vague (i.e., the toy only plays half of a song and half of the lights do not work) to elicit as many troubleshooting strategies and thinking processes as possible without the intensive labor of actually solving a project. Thus, the failure artifact scenarios allow for a means of inquiring into students' troubleshooting thought processes at multiple time points without physically and time-intensive scenarios like pre-created buggy physical computing projects (Lui et al., 2024) or escape rooms (Tilman & Romeike, 2020). Failure artifact scenarios complement these other means of analyzing students' troubleshooting thought processes. Here, it is worth noting that the instrument did not assess students actual troubleshooting processes but their thinking and explanations of how they would go about troubleshooting. We acknowledge that students might excel in discussing troubleshooting but not in practicing troubleshooting (Collins & Evans, 2019). Therefore, we suggest using instruments like this one in conjunction with other observational methods, including students' own designed projects or portfolios that document changes made and bugs solved (Fields, Lui et al., 2021). At the same time, being able to explain how to troubleshoot a problem and articulate plausible strategies is important evidence of students' development of computational communication—the ability to understand and communicate computational thought processes, procedures, and results to others—a key computing practice.

Finally, as a research tool, the failure artifact scenarios may provide a starting point for others to assess and study students' understanding of troubleshooting in other domains of physical computing, such as robotics. The instrument could also be used to assess students' troubleshooting thinking processes longitudinally, through clinical interviews at different points of the school year or across their high school trajectory. The scenarios could also be applied in a quasi-experimental design to see if one intervention is more successful at supporting students' learning of troubleshooting than a control group. For this, it might be helpful to allow written responses to failure artifact scenarios (e.g., Simon et al., 2008) or to create survey constructs, exploring in pre- and post-measures whether students identify potential multiple causes of problems across multiple domains and with increasing levels of specificity in steps, processes, and solution paths.

# Conclusions

In this paper, we explored the design and use of failure artifact scenarios to document changes in students' understanding and thinking of troubleshooting of physical computing projects. Situated in the context of electronic textiles, the analysis of pre- and post-clinical interviews illustrated changes in students' locating sources of bugs across software and hardware, often with increasing specificity, attention to multiple domains, and iterative testing. Inviting novice learners to think about failure rather than functioning artifacts may help other researchers and teachers gain insights into students' understanding of the complex, multimodal process of troubleshooting.

# Acknowledgements

With regards to Katherine Gregory for support in data analysis. This work was supported by a grant from the National Science Foundation to Yasmin Kafai and Mike Eisenberg (#1742140/#1742081). Any opinions, findings, conclusions, or recommendations expressed in this paper are those of the authors and do not necessarily reflect the views of NSF, the University of Pennsylvania, or Utah State University.

# Appendix 1.

## Pre-Unit Interview Protocol

### Scenario 1.

Interviewer: I'm gonna show you two projects that are not working and I want to ask you to help me figure out how to fix them. I know you have not started the e-textiles unit, but anything you tell me will be helpful. Tell me what you are thinking and to share any ideas you may have so that I can try them later. Here is an e-textiles project that a student in another class is working on. It's not working as intended. [Interviewer shows image (see Figure 8) by sharing a document that the student can annotate. Interviewer reads the notes below the images.]

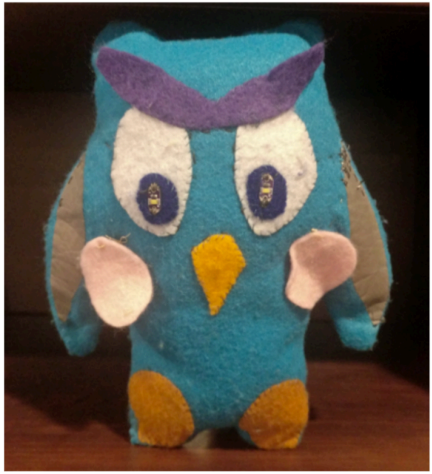

The lights on this toy bird's eyes should light up and the bird should sing a melody whenever someone presses both its wings (silver patches on either side.)

But, one of the lights isn't turning on and the bird sings the melody only half way through whenever someone presses both its wings (silver patches on either side).

Figure 8. Image with description of failure artifact scenario 1 presented to students in the pre-interview.

Interviewer: What would be your instructions for that student to fix it?
   a. Feel free to annotate the picture , if it's helpful to you, but you don't have to.
   b. [follow-up as needed] Remember, we can ask the student to unsew some stitches and open up the toy too.
   c. [follow-up as needed] Remember, we can direct the student to look into the code as well.
   d. [follow-up as needed] Why did you say that?

## Scenario 2.

Interviewer: Here is my Captain America Shield project [show the actual project and share "teacher-approved" circuit drawing (see Figure 9)]. I also have a simulation of the project so that you can interact with it. Here is how it is supposed to behave [demo how it works! https://scratch.mit.edu/projects/479329018/fullscreen/ ]:
   e. if both buttons are pushed, the outer LEDs (on white ring) turn on, while the inner LEDs (on red ring) do a chase pattern;
   f. if button1 is pushed and button2 is not pushed, the outer LEDS (on the white ring) are off, and the inner LEDs (on the red ring) blink;
   g. if button1 is not pushed and button2 is pushed, the outer LEDs (on the white ring) blink, and the inner LEDS (on the red ring) are off;
   h. if neither button is pushed, then all LEDs are off.

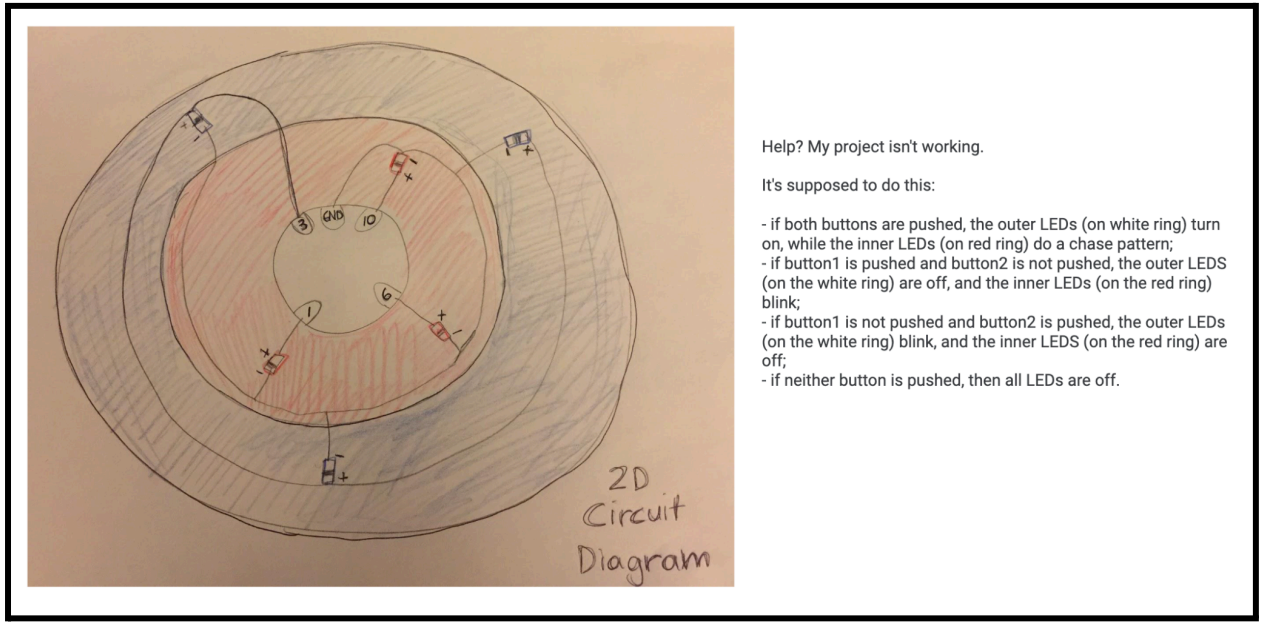

Figure 9. Image of circuit diagram with description of failure artifact scenario 2 presented to students in the pre-interview.

Inteviewer shares the link [htttps://scratch.mit.edu/projects/479329018/fullscreen/] with the student and requests that they share their screen.

Inteviewer: This project isn't working. Can you tell me how to fix it? I can write down steps to do when I go back to my workshop. Talk about everything you are thinking as you look at this project.
  a. [follow-up as needed] What do you think could be the causes for each of these issues?
  b. [follow-up as needed] How would you fix them?
  c. Why did you say that?
  d. When I leave, what should I do?

# Post-Unit Interview Protocol

## Scenario 1.

Interviewer: Here is an e-textiles project that a student in another class is working on.. Its not working as intended. [Interviewer shows image (see Figure 10) by sharing a document that the student can annotate. Interviewer reads the notes below the images.]
What would be your instructions for that student to fix it?
  a. Feel free to annotate it. Feel to mark up the picture too.
  b. [follow-up as needed] Remember, we can ask the student to unsew some stitches and open up the toy too.

c. [follow-up as needed] Remember, we can direct the student to look into the code as well.

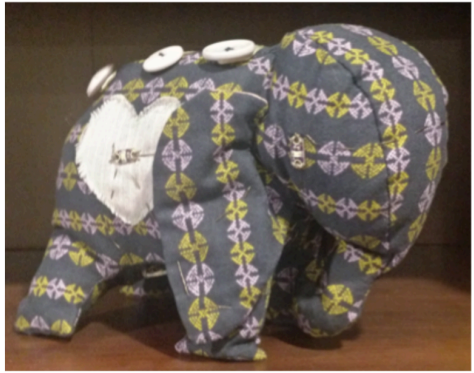

The lights on the side (heart shape) of this elephant should light up and its trunk should move up and down whenever someone presses both its ears.

But, one of the lights on its side flickers and its trunk moves only halfway up whenever someone presses both its ears.

Figure 10. Image with description of failure artifact scenario 1 presented to students in the post-interview.

## Scenario 2.

Interviewer: Here is my Bear project [show the actual project and share "teacher-approved" circuit drawing (see Figure 11) ]. I also have a simulation of the project so that you can interact with it. Here is how it is supposed to behave [demo how it works! https://scratch.mit.edu/projects/517391179/fullscreen/]. Here is how it is supposed to behave:

   a. If both buttons are pressed, all lights are on
   b. If no buttons are pressed, both the eye lights blink and the cheek lights blink.
   c. When button 1 is pressed, only the eye lights blink fast.
   d. When button 2 is pressed, only the cheek lights blink fast.

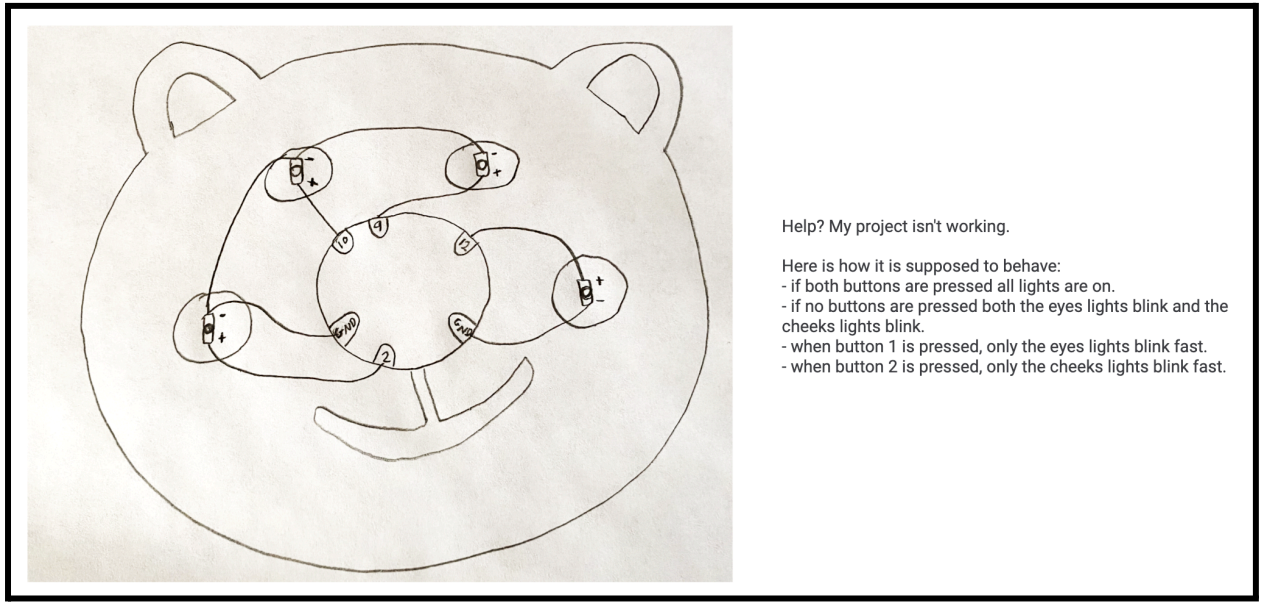

Figure 11. Image of circuit diagram with description of failure artifact scenario 2 presented to students in the post-interview.

Inteviewer shares the link [https://scratch.mit.edu/projects/517391179/fullscreen] with the student and requests that they share their screen.

Interviewer: This project isn't working. Can you tell me how to fix it? I can write down steps to do when I go back to my workshop. Talk about everything you are thinking as you look at this project.

      e. [follow-up as needed] What do you think could be the causes for each of these issues?
      f. [follow-up as needed] How would you fix them?
      g. Why did you say that?
      h. When I leave, what should I do?

# Appendix 2.

| **Articulating potential bugs** | coding | Student articulates an undefined coding bug. They articulate coding as the relevant domain but do not articulate a bug | Student articulates a semi-defined coding bug. They articulate coding as the relevant domain and articulate either a bug location or a specific | Student articulates a well-defined coding bug. They articulate coding as the relevant domain, articulate a bug location, and articulate a specific error. |

|  |  | location or a specific error. E.g., "I guess since like the, the head of the elephant has to move up, I guess the… they have to learn, like, what to code." | error, but not both. E.g., "they probably didn't, like, put it in the code where pressing both of them equals, like, the movement and the lights turning on." | E.g., "I think you should check for, like, the labeling of the code. Like make sure you label exactly where the light goes." |
|---|---|---|---|---|
|  |  | Codes (p = prompted, np = not prompted): Id_code.undefined.p Id_code.undefined.np | Codes (p = prompted, np = not prompted): Id_code.semi-defined.p Id_code.semi-defined.np | Codes (p = prompted, np = not prompted): Id_code.defined.p Id_code.defined.np |
|  | circuit | Student identifies an undefined crafting/circuitry bug. They articulate crafting/circuitry as the relevant domain but do not articulate a specific bug location or a specific error. E.g., "something is wrong with the cables inside of the bird" | Student identifies a semi-defined crafting/circuitry bug. They articulate crafting/circuitry as the relevant domain and articulate either a specific bug location or a specific error, but not both. E.g., "Maybe the thread is not sewed on tight." | Student identifies a well-defined crafting/circuitry bug. They articulate crafting/circuitry as the relevant domain, articulate a specific bug location, and articulate a specific error. E.g., "Is it because [the thread]'s not connected to GND?" |
|  |  | Code: Id_cc.undefined | Code: Id_cc.semi-defined | Code: Id_cc.defined |
|  | unknown | Student identifies a bug but does not explicitly indicate its domain. E.g., "I try to tweak, like, the music thing so that it plays all the way through when the wings are pressed. " |||
|  |  | Code: Id_unknown |||
| **Multiple causes** | Student considers that one specific bug/problem may be caused by more than one issue <u>within a single domain</u>. E.g., "you can check the inside of the elephant to see if, to see if there, if, if the wires or, or yeah. Or the wires or lights are connected correctly. Could one of the lights be burned out?" E.g., I guess, um, in the programming, they were… they had to program like the, the button, like the buttons. They have to press, like, the button in one storage or like the void loop pattern. I guess they, um, they, uh, didn't, like, put in the code where like the light has to stay on. || Student considers that one specific bug may be caused by more than one issue <u>in different domains.</u> E.g., "Maybe connect this one [pointing with cursor to LED] to this one [pointing with cursor to LED]... maybe this one isn't like connecting well… oh, or maybe in the code. They probably forgot it [the LED]." ||

|  | Codes (c=code, cc=circuit/craft):<br>Causes.single.c<br>Causes.single.cc | Codes:<br>Causes.multi | |
|---|---|---|---|
| **Proposed next steps for fixing buggy projects** | Student provides vague next steps. Within the next steps, they articulate a relevant domain but do not specify a location (e.g., component or section of the code).<br><br>E.g., "just double check [the code] basically and test it to see what's wrong." | Student provides concrete next steps for fixing a bug. Within the next steps, the student specifies a relevant domain and identifies a specific location (e.g., component or section of the code).<br><br>E.g., "Maybe connect this one to this one. [points with mouse from right "eye" to left "eye"] because it, maybe this one [pointing to thread and left "eye" LED] isn't like connecting well" | |
|  | Codes (unknown=domain unidentified, c=code, cc=circuit/craft)<br>NextSteps.vague.unknown<br>NextSteps.vague.c<br>NextSteps.vague.cc | Codes (c=code, cc=circuit/craft)<br>NextSteps.concrete.c<br>NextSteps.concrete.cc | |
| **Debugging Process** | Student articulates **a vague sequence** of steps for identifying bugs or/and solving buggy projects without acknowledging the domain(s) of the bug.<br><br>E.g.,"Um, you could check, like, I don't know how it works, but the ears, like when you press 'em, if everything with that is right." | Student articulates **an undefined sequence** of steps for identifying bugs or/and solving buggy projects. They neither identify the process nor the location of the process (how and where).<br><br>E.g.,"Maybe you should check if like the lights on the right thing, like the positive, the negative sides or the light thing, the little LED lights or whatever"<br><br>E.g., "Maybe test out the circuit before, um, actually like sewing it on."<br><br>E.g., "You could recheck the code and rewrite it" | Student articulates **a semi-defined sequence** of steps for identifying bugs or/and solving buggy projects. They either identify the process or the location of the process.<br><br>E.g., "Um, I say, uh, like check the wing, cause in the wing, uh, to like press the, the, uh, lights, like I'll wing to see if anything was wrong with it."<br><br>E.g., "I feel like it's the same thing to, like, double check work back from the beginning, start from the beginning and work your way up again, and then you'll probably see your mistake." | Student articulates **a defined sequence** of steps for identifying bugs or/and solving buggy projects. They have identified both the process and the location of the process.<br><br>E.g., "Make sure the buttons are connected, like to press on the lights, like make sure those are, like, together. I don't know."<br><br>E.g., "I would tell the student to go back to its coding and, and then if it's, uh, check through its, uh, comments that it put on the code and to see what you did" |
|  | Codes (c=code, cc=circuit/craft, multi) | Codes (c=code, cc=circuit/craft, multi)<br>Process.c.undefined<br>Process.cc.undefined<br>Process.multi.undefined | Codes (c=code, cc=circuit/craft, multi)<br>Process.c.semi-defined<br>Process.cc.semi-defined<br>Process.multi.semi-defined | Codes (c=code, cc=circuit/craft, multi)<br>Process.c.defined<br>Process.cc.defined<br>Process.multi.defined |

|  | Process.unknown.undefined |  |  |  |
|---|---|---|---|---|
| **Personal experience** | Student makes a reference to personal experience in relation to debugging.<br><br>E.g., "Maybe it's like, uh, in the Christmas tree, if one goes out, then they all go out." | | | |
| | Codes (t=teacher, p=projects, o=other):<br>Experience.teacher<br>Experience.projects<br>Experience.other | | | |
| **Testing project** | Students test all four conditions andarticulate the problems that come up (Only applicable for question 2)<br>Codes (p = prompted, np = not prompted):<br>Testing.p<br>Testing.np | | | |